# Thirty-Year Period in Secular Variation of the Main Geomagnetic field

**Wen-Yao Xu**
(Institute of Geology and Geophysics, Chinese Academy of Sciences, Beijing, China)
**Henri-Claude Nataf**
(Laboratoire de Geophysique Interne et Tectonophysique, Observatoire des Sciences de l'Univers de Grenoble, France)
**Zi-Gang Wei and Ai-Min Du**
(Institute of Geology and Geophysics, Chinese Academy of Sciences, Beijing, China)

**Abstract**

During the centennial period from 1900 to 2000, the globally averaged unsigned annual rate of the main geomagnetic field experienced a three-episode variation. The maximum annual rates occurred respectively around 1910-1920, 1940-1950, and 1970-1980, showing a 30-year period. In addition, the rising phase in each episode is much shorter than declining phase. The governing factor of this periodic variation is non-dipole field, instead of the dipole field, although the secular variation of the magnetic energy concerned with the dipole is dominative over all other multiples.

**1. Introduction**

The main geomagnetic field is subject to changes in its direction, strength and spatial pattern on various timescales, known as secular variation (SV). Among the verity of features of the SV, periodic or cyclic features have long attracted studies. Different periods, ranging from years to millennia and even millions of years, have been revealed by the studies on paleomagnetism, archeomagnetism and historical data analysis. Polarity reversal associated with the strength change has long, but irregular period, from tens of thousands to millions of years (Merrill and McFadden,1999, Guyodo and Valet,1999,Valet 2003). The dipole axis rotates around the geographic axis at a rate of about $0.05°$/year, showing an 7000-year period. The non-dipole field drifts westward at a rate of $0.2°$/year, and will take 1800 years for a complete round (Malin and Saunders 1973, Langel 1987, Wei and Xu, 2000, 2001,2002,2003). The variations of declination and inclination at London imply a period of some 600 years (Chapman and Bartels, 1940). Long records have been analyzed by using various spectrum techniques, such as Fourier analysis, wavelet analysis, Maximum Entropy Spectrum analysis (MESA), showing a well-known "century period" from 60 years to 100 years (Wang et al, 1982 ). A typical short-period variation in SV is geomagnetic jerk, which repeatedly occurs about every 10 years in twenty century (). In addition, archeomagnetic jerks are also examined (Gallet et al 2003,).

Physical aspects of the periodicity in SV have been studied for a few decades. As widely accepted, the main geomagnetic field and its secular variations originate from fluid flow in the Earth's core. The fluid flow at the surface of the Earth's core can be deduced from the geomagnetic data observed at the surface (Gubbins, 1982, Bloxham and Jackson, 1991). The knowledge on the fluid flow within the core, however, mainly relies on numerical MHD simulations (Glatzmair and Roberts, 1995, Kuang and Bloxham, 1997, Kono and Roberts, 2002).

Polarity reversals depend on changes of the fluid flow in the core (Glatzmaier and Roberts, 1995), although the Earth's mantle controls, on a great extent, the frequency of geomagnetic reversals (Glatzmaier et al, 1999). In order to understanding westward drift in the main field, Jault et al (1988) examined the correlation of westward drift with exchanges of angular momentum between core and mantle, and Hide (1966) suggested a propagating magnetic mode of MHD wave in the outer core. As for the so-called century period, IGRF models supply a proper database. Using the method of Natural Orthogonal Components (NOC), Xu and Sun (1998) analyzed the relationship between the variation periods of the main field and its special configuration, showing an interesting association of the eigenmodes (principal components) with special inherent periods. Bellager et al (2001) correlated geomagnetic jerks with Changler wobble. Bloxham et al (2002) show that geomagnetic jerks can be explained by the combination of a steady flow and a simple time-varying axisymmetric, equatorial symmetric toroidal zonal flow.

In this paper the 9-th generation of IGRF (IAGA Division 5 Working Group 8, 2000, 2003) is used to study periodical characteristics in the secular variation of the main geomagnetic field for recent 100 years.

## 2. Secular variation of the magnitudes of SV-field

The main geomagnetic field is conventionally divided into dipole part (briefly, DP) and non-dipole part (ND). The energy density (also often called the "power") of the geomagnetic field of degree $n$ on the Earth's surface may be written in terms of Gauss coefficients $g_n^m$ and $h_n^m$ as (Lowes, 1974, Langel 1987)

$$R_n = (n+1)\sum_{m=0}^{n}[(g_n^m)^2 + (h_n^m)^2] \qquad (1)$$

The left panel of Figure 1 shows the dependence of $R_n$ on the harmonic degree n, usually referred to as the geomagnetic spectrum. It is noted that in the main field the power $R_1$ of the dipole field ($n=1$) obviously dominant over other multipoles $R_n$ ($n=2, 3,\ldots$), making the geomagnetic field pattern look like a dipole field.

The secular variation (SV) of the main field is the first time-derivative of the field, usually represented by annual rate of the field. From formula (1) the annual rate of the field power is deduced and referred to as the "SV spectrum"

$$\dot{R}_n = 2(n+1)\sum_{m=0}^{n}[g_n^m \dot{g}_n^m + h_n^m \dot{h}_n^m] \qquad (2)$$

where $\dot{g}_n^m$ and $\dot{h}_n^m$ are the first time-derivative (or annual rate) of the Gauss coefficients $g_n^m$ and $h_n^m$. Since the Gauss coefficients of the dipole field, mainly $g_1^0$, are much greater than the coefficients of the high-degree multipoles, even a small annual rate $\dot{g}_1^0$ will leads to a large $\dot{R}_1$.

Consequently, it is expected that the annual rate of the dipole field power, $\dot{R}_1$, is dominant in the SV spectrum $\dot{R}_n$, as shown in the middle and right panels of Fig 1. This factor may mislead to a incorrect conclusion: the secular variation of the dipole field also dominants the SV-field pattern (or spatial distribution of the annual rate).

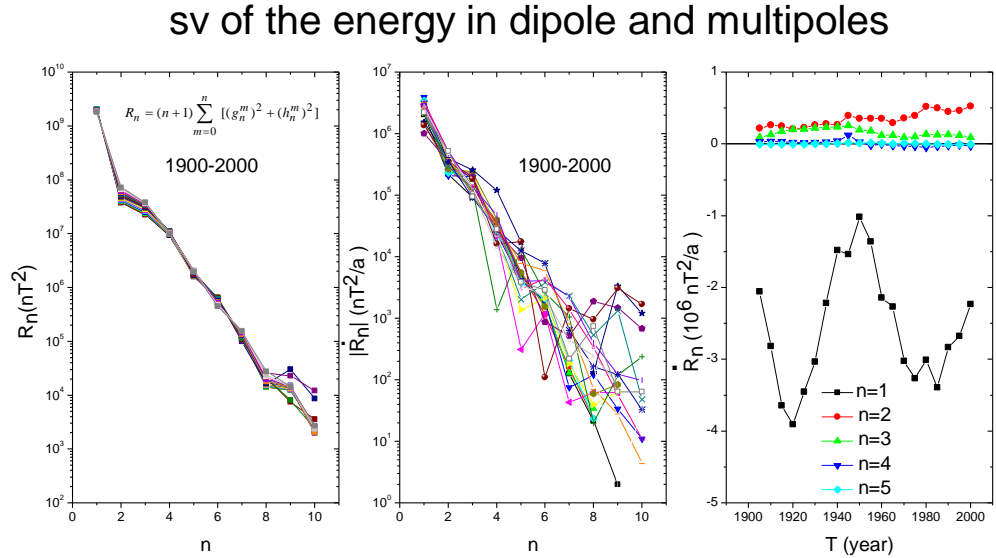

Fig 1. The geomagnetic spectrum (left), the SV spectrum (middle), and the annual rates of the dipole and multipoles (right).

In fact, the SV-field pattern is dependent on $\dot{g}_n^m$ and $\dot{h}_n^m$, instead of power $R_n$. Using two successive IGRF models (IAGA Division 5 Working Group 8, 2000, 2003) for epochs $t_1$ and $t_2$, one can obtain the average annul rate of secular variation of the main field during this time span. Fig. 2 shows the secular variation field (SV-field) of Z component calculated in this way for several 5-year intervals. It is noted in the figure that the most prominent features of the SV-field are several local regions of secular variations, instead of global decrease, as the dipole moment declines suggests. This fact implies the important role of the non-dipole field in the SV-field.

IGRF-SV1900-2000 (total intensity F, nT/a)

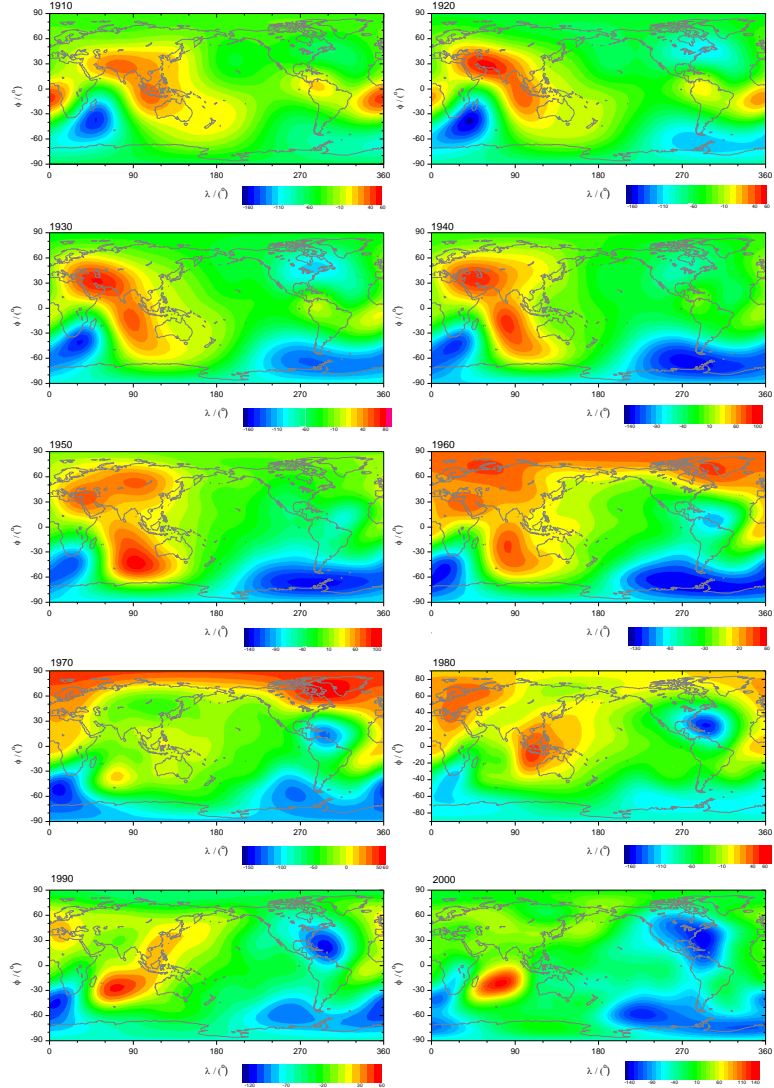

Fig 2.   Secular variation field (SV-field) of Z component.

In this paper we will focus on global magnitude of the SV-field, instead of its detailed spatial distribution.   In order to describe the overall magnitude of the global SV, a proper measure is needed.   Bondi and Gold (1950) elucidated a quantity, pole-strength, as an useful measure of the strength of an internally generated magnetic field

$$= | \quad | \qquad (3)$$

where $S$ is a closed surface surrounding the magnetic sources, such as the surface of a planet, is the outward unit vector of the surface, **B** the magnetic field at the surface.   This quantity, later called "unsigned magnetic flux", is widely used in geomagnetism (Roberts and Scott, 1965,

Hide, 1978, Hide and Malin, 1981, Li Kai et al., 1986, Benton et al., 1987, Voorhies and Benton, 1982, Xu, 2001). Since the contributions from both the dipole and the multipoles are included, the pole-strength is an adequate measure for describing the global features of the field, in comparison with nother parameters, such as the dipole moment, the maximum field strength, and average value in a certain area or whole glob.

Following this idea, we define a globally averaged unsigned annual rate of Z component as follows

$$\bar{\dot{Z}} = \frac{1}{S} \int_0^\pi \int_0^{2\pi} |\dot{Z}| r^2 \sin\theta \, d\lambda \, d\theta \qquad (4)$$

where $\theta$ and $\lambda$ are colatitude and longitude, respectively. $\dot{Z}$ represents the annual rate of Z component at location $(\theta, \lambda)$. Similarly, we have $\bar{\dot{X}}$, $\bar{\dot{Y}}$, $\bar{\dot{H}}$ and $\bar{\dot{F}}$ for northward, eastward and horizontal components, as well as total intensity.

Fig.3 shows the variations of the globally averaged unsigned annual rates of 5 elements *X, Y, Z, H* and *F* by using the 9-generation of IGRF, which includes 21 main field models for every 5 years from 1900 to 2000 and one secular variation model for 2000-2005. It is interesting to note that the annual rates of these components consistently show a fairly regular oscillation variation. During the centennial period from 1900 to 2000, the secular variation of the main geomagnetic field experienced a three-episode variation. The maximum annual rates of these elements consistently occurred around 1910-1920, 1940-1950, and 1970-1980, showing a 30-year period variation. In addition, the rising phase in each episode is much shorter than declining phase. The detail information of the annual rates are listed in Table 1.

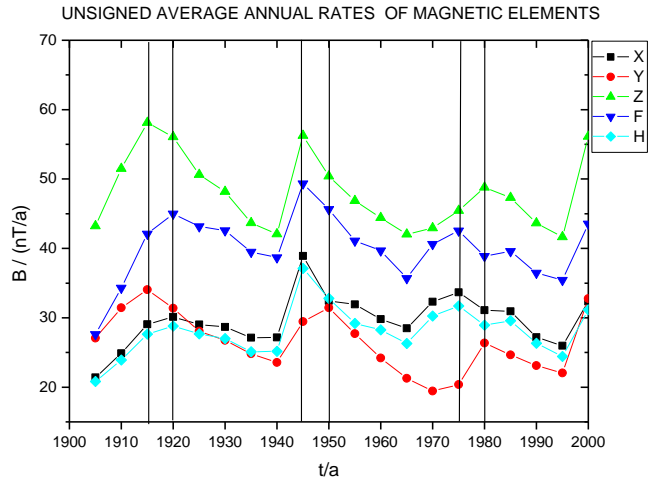

Fig 3. Variations of the globally averaged unsigned annual rates for *X, Y, Z, H,* and *F*

Table 1. Characteristics of the main field secular variation for 1900-2000

| 1 | 2 | 3 | 4 | 5 | 6 | 7 | 8 |
|---|---|---|---|---|---|---|---|
|   | *mean* (nT/a) | *max* (nT/a) | *min* (nT/a) | *range* (nT/a) | *1-st peak* | *2-nd peak* | *3-rd peak* |

|   |   |   |   |   | $t_{max1}$ | $t_{max2}$ | $t_{max3}$ |
|---|---|---|---|---|---|---|---|
| $\dot{X}$ | 29.6 | 38.9 | 21.4 | 17.5 | 1915-1920 | 1940-1945 | 1970-1975 |
| $\dot{Y}$ | 26.5 | 34.0 | 19.4 | 14.6 | 1910-1915 | 1945-1950 | 1975-1980 |
| $\dot{Z}$ | 48.0 | 58.1 | 41.7 | 16.5 | 1910-1915 | 1940-1945 | 1975-1980 |
| $\dot{H}$ | 28.1 | 37.1 | 20.8 | 16.3 | 1915-1920 | 1940-1945 | 1970-1975 |
| $\dot{F}$ | 40.1 | 49.3 | 27.6 | 21.7 | 1915-1920 | 1940-1945 | 1970-1975 |
| **Average** |   |   |   |   | 1910-1920 | 1940-1950 | 1970-1980 |

### 3. Origin of the periodic variation in the SV-field

The most prominent feature in the secular variation is steady decrease of the dipole moment, but with different rate. Fig. 4 shows this general decrease tendency of the dipole moment $M_{DP}$ and its annual rate $\dot{M}_{DP}$. It is interesting to note that the annual rate (magnitude) of the dipole moment reaches maximum around 1930 and 1980, while there is a low rate around 1950. If the 1930' and 1980's peaks in Fig. 2 might be attributed to the dipole field variation, the 1950' peak is hard to be explained by same reason. In fact, even the 1930' and 1980's peaks are mainly caused by the non-dipole field.

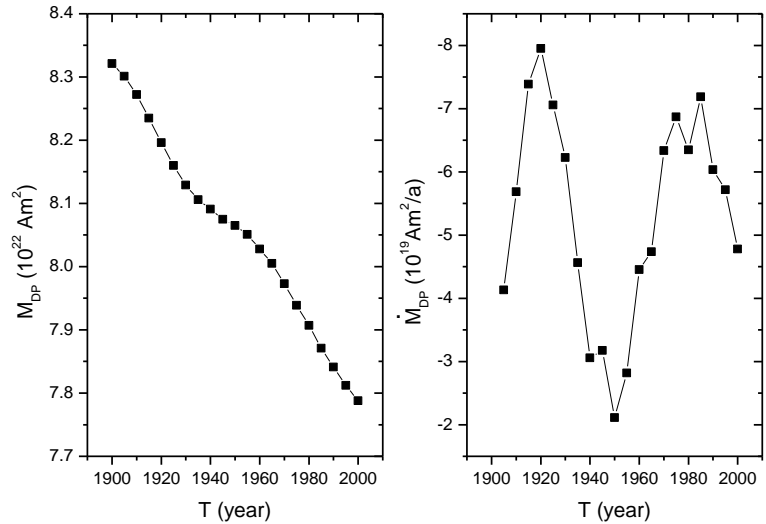

Fig 4. Variations of the dipole moment of the geomagnetic field (left) and its annual rate (right).

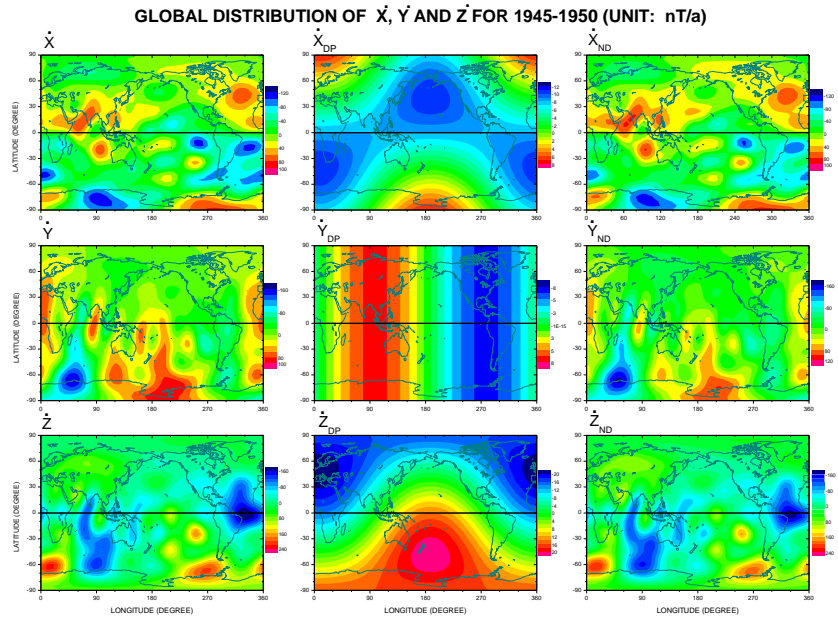

Fig5. Global distributions of the total SV-field for 1990-1995 (left hand column), the contributions from the dipole field (middle column), and the contributions from the non-dipole field (right hand column). From top to bottom: $\dot{X}$, $\dot{Y}$ and $\dot{Z}$

In order to confirm the contributions of the dipole and non-dipole fields to SV-field, the global distributions of SV-field are calculated for the dipole, non-dipole and total SV-fields As an example, Fig. 5 illustrates the global distributions of the total SV-field, $\dot{X}$, $\dot{Y}$ and $\dot{Z}$ (left hand column), the contributions from the dipole field (middle column) and from the non-dipole field (right hand column). It is obvious that the overall patterns of secular variations are governed by the non-dipole field. Fig 6 shows the unsigned average annual rates of the dipole and non-dipole fields, compared with the total rates. In general, $\bar{\dot{Z}}_{ND}$ is about twice of $\bar{\dot{Z}}_{DP}$. The rate $\bar{\dot{Z}}_{DP}$ for 1950 is as low as 10 nT/a, while the rate $\bar{\dot{Z}}_{ND}$ is about 55 nT/a, dominant the total rate..

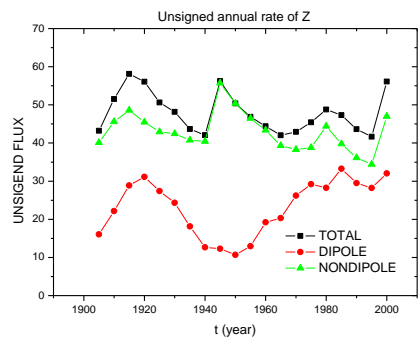

Fig 6. Globally averaged unsigned annual rates of the dipole, non-dipole and total field for Z component.

**4. Summary**

   (1) During the centennial period from 1900 to 2000, the secular variation（the first time-derivative）of the main geomagnetic field experienced a three-episode variation.  The maximum annual rates occurred respectively around 1910-1920, 1940-1950, and 1970-1980, showing a 30-year period.  In addition, the rising phase in each episode is much shorter than declining phase.

   (2) The governing factor of this periodic variation is non-dipole field, instead of the dipole field, although the secular variation of the magnetic energy concerned with the dipole is dominant over other multiples.

**Acknowledgement**   This study is supported by the fund from Chinese Committee of Natural Science Fundation (40436016).

Figure captions

Fig 1.　The geomagnetic spectrum (left), the SV spectrum (middle), and the annual rates of the dipole and multipoles (right).

Fig 2. Secular variation field (SV-field) of Z component.

Fig 3. Variations of the globally averaged unsigned annual rates for *X, Y, Z, H,* and *F*

Fig 4. Variations of the dipole moment of the geomagnetic field (left) and its annual rate (right).

Fig 5. Global distributions of the total SV-field for 1990-1995 (left hand column), the contributions from the dipole field (middle column), and the contributions from the non-dipole field (right hand column). From top to bottom: $\dot{X}$, $\dot{Y}$ and $\dot{Z}$

Fig 6. Globally averaged unsigned annual rates of the dipole, non-dipole and total field for Z component.